\documentclass[twocolumn,10pt]{article}
\usepackage{times}
\usepackage{latex8}
\usepackage{url}
\usepackage{epsfig}
\usepackage{epstopdf} 
\usepackage{amssymb}
\usepackage{amsmath}
\newcommand{\newln}{\\&\quad\quad{}}
\begin{document}

\title{Accurate Local Estimation of Geo-Coordinates for Social Media Posts}
\author{Derek Doran and Swapna S. Gokhale\\
Dept. of Computer Science \& Eng. \\
University of Connecticut\\
Storrs, CT, 06269\\
\{derek.doran,ssg\}@engr.uconn.edu
\and
Aldo Dagnino\\
Industrial Software Systems\\
ABB Corporate Research\\
Raleigh, NC, 27606\\
aldo.dagnino@us.abb.com
}

\maketitle

\begin{abstract}
Associating geo-coordinates with the content of social media 
posts can enhance many existing applications
and services and enable a host of new ones. Unfortunately, a 
majority of social media
posts are not tagged with geo-coordinates. Even when location data
is available, it may be inaccurate,
very broad or sometimes fictitious. Contemporary
location estimation approaches based on analyzing the content of these posts 
can identify only broad areas such as a city, which limits their usefulness. 
To address these shortcomings, this paper 
proposes a methodology to narrowly estimate the
geo-coordinates of social media posts with high accuracy. 
The methodology relies solely on the content of these posts 
and prior knowledge of the wide geographical region
from where the posts originate. An ensemble of language models, 
which are smoothed over non-overlapping sub-regions of a wider region, 
lie at the heart of the methodology. Experimental evaluation 
using a corpus of over half a million tweets from New York City 
shows that the approach, on an average, estimates locations of tweets to 
within just $2.15$km of their actual positions.
\end{abstract}


\section{Introduction}
The information shared by users over Online Social Networks 
(OSNs) such as Facebook and Twitter offer unique insights into their
thoughts, emotions, and opinions. The richness of these posts
has motivated numerous organizations to harvest the content embedded
within them in support of value-added services. Associating geographic 
locations with the information extracted from these posts can offer
both theoretical and practical benefits. Theoretically, 
this association can facilitate sociological studies to examine 
how online behaviors, relationships, and interactions are influenced 
by their offline socio-spatial counterparts~\cite{scellato11}. 
Practically, the linking of location to content can enhance 
existing services such as location-based advertising~\cite{kolmel02}
and disaster response~\cite{yin12} and conceive novel ones.

Social media posts may be tagged with location information in two
ways. First, users may choose to automatically tag their posts 
shared via GPS-enabled mobile devices. Second, many OSNs 
allow users to include their current location~\cite{sadilek12} through fields 
such as ``location'' or ``from'' in their social media profiles. 
If users diligently and authentically use one of these two methods,
then accurate location information can be extracted easily.  
However, currently, a vast majority of users do not enable tagging of
their mobile posts~\cite{cheng10} and choose not to include their
locations in their profiles, perhaps for privacy reasons. Some 
users who do populate this field may specify it broadly in terms of 
a state or a country, while some may intentionally
provide inaccurate or fictitious positions~\cite{hecht11}.
Thus, in practice, only a small percentage of social media posts 
are accompanied by rich and accurate location data. To alleviate this 
shortcoming, contemporary approaches that need the location of posts 
estimate it by analyzing their content. These approaches, however,
estimate broad regions of the order of a city, or 
location ``types'' such as restaurants, offices, homes, or 
stores~\cite{dalvi12,li12}. Finally, a few efforts that try to estimate the 
actual positions or geo-coordinates of social media posts are 
accurate within a radius of 80-100 km~\cite{cheng10,mahmud12,cho11},
essentially identifying only broad regions. 

Once the broad region from where a social media post has originated 
is identified either through tagging, by finding keywords 
corresponding to 
famous landmarks and interesting events, or by using an aforementioned 
contemporary approach, pinning it down narrowly to within a 
small radius around its actual geo-coordinates 
may add significant value. For example, such local geo-tagging can 
shed light on how people within a neighborhood think similarly 
as they are exposed to common events, and participate in richer and meaningful 
offline friendships~\cite{mok10}. Identifying localities 
can also provide  more accurate information on an event or a disaster 
which can be highly beneficial to first responders~\cite{yin12}. 
Law enforcement can also use such fine 
geo-tagging to approximate the location of a suspect, who is known to 
be present in a town or a city. Finally, it may be feasible to
identify the geo-coordinates of a post with high accuracy once its broad
region is known because this prior knowledge limits the range of 
possible positions. 

In this paper, we present a methodology to accurately estimate the 
geo-coordinates of a social media post based on its content, once 
the broad region from where it originates is known.  
The methodology consists of partitioning the broad region into a grid of 
non-overlapping sub-regions, building probabilistic language models over 
each sub-region, and then applying geo-smoothing to improve the accuracy of 
the location estimates. We train and evaluate the language models over a 
corpus of tweets collected across downtown and midtown Manhattan, 
and find that the approach, on an average, pinpoints the 
positions of tweets to within $2.15$km, or just 4\% of the size of the 
total region from which they are known to originate.

This paper is organized as follows. Section~\ref{sec:method} describes
our estimation methodology. Section~\ref{sec:results} presents 
experimental evaluation. Related work
is compared in Section~\ref{sec:rr}. Conclusions and future work are
offered in Section~\ref{sec:conc}. 

\section{Estimation Methodology}
\label{sec:method}
In this section, we describe the two steps in 
the methodology to estimate geo-coordinates.

\subsection{Building Language Models}
The topics, thoughts, words, and expressions embedded within social
media posts are influenced by the inherent properties and circumstances 
of the locality from where users share these posts. 
For example, people may share their opinions of a restaurant while
seated there. They may also share about an accident or a 
noteworthy public event as it occurs. The culture
and social norms of a local area may also modulate these posts. 
As an example, posts from Little Italy in New York City may be 
pre-dominantly influenced by Italian norms and culture, while those 
from Times Square may instead overwhelmingly share the excitement
of visiting the city. 
Thus, both the language and the content of social media posts shared from 
different, smaller sub-regions within a broad region will be varied. 

To expose these local variations, we partition a broad region of 
interest $\mathcal{L}$ into a collection of equally sized, non-overlapping 
sub-regions $\ell_i$, which are defined by a $g \times g$ grid.
We then build an ensemble of models; one per sub-region to represent the 
language of its posts. This ensemble is inspired by recent approaches 
including our own~\cite{snaa13,seke13}, that have demonstrated 
its promise in capturing the linguistic variations in the content of 
social media posts. A language model defines a probability distribution over 
$n$-grams, where an $n$-gram is an ordered sequence of $n$ 
words $(w_1,....,w_n)$. The maximum likelihood estimate of 
an $n$-gram, computed over a corpus of posts
within $\ell_i \in \mathcal{L}$, is given by~\cite{bahl83}: 
$$P_{\ell_i}(w_1,...,w_n) = \frac{c_{\ell_i}(w_1,...w_n)}{c_{\ell_i}(w_1,...,w_{n-1})}$$
where $c(.)$ is the number of times the sequence appears in the posts. 
The probability that a sub-region 
generates a phrase $T = (w_1,...,w_k)$ is computed as the 
product of the probabilities of the $n$-grams that comprise $T$:
\begin{equation}
P(T |\ell_i) = \prod_{j=1}^{k-n+1} P_\ell(w_{j},w_{j+1},...,w_{j+n-1}) \nonumber
\label{eq:one}
\end{equation}

Contextual information increases with $n$ because 
longer sequences of words can be considered. However, 
because social media posts are short, specific long word sequences 
appear with low
frequency, and hence, prevalent approaches use only unigrams to model 
these posts. Although unigrams or $1$-grams model 
the distinct vocabulary, independent of the 
order of words~\cite{chandra11}, they lack the ability to capture
context within a language. For example, a unigram model
trained over ``going to work'' can represent how one
discusses the concept of ``work'', and the action of 
``going'', but cannot associate the concept with the action. 
Language models trained over bigrams 
``going to'' and ``to work'', however, can capture additional context
of going somewhere, and applying an action or a verb to the concept 
of ``work''. We limit to bigrams although higher order models can
capture even more details, because estimating higher order models 
may be inaccurate using a corpus of social media posts that are typically 
short but refer to a broad variety of topics. 

To improve the accuracy of the language models, we interpolate 
the probability of a bigram with the probability of the unigram that 
completes it. This interpolation compensates for the low count 
of a bigram by incorporating the expected higher count of the unigram 
that completes it. For example, 
if the unigram ``driving'' is used frequently in a training
corpus, we should expect that bigrams completed by this word (e.g. ``love
driving'') are more likely to be seen even if the bigram does not appear often.
Thus, for a sub-region 
$\ell_i$, the probability of observing the bigram $(w_{j-1},w_{j})$ is 
given as: 
$$P_{\ell_i}(w_{j-1},w_j) = 
\lambda_1 \frac{c(w_{j-1},w_j)}{c(w_{j-1})} + 
\lambda_2 \frac{c(w_j)}{|W(\ell_i)|}$$ 
where $\lambda_1 + \lambda_2 = 1$,
$|W(\ell_i)|$ is the number of distinct words in all posts in
$\ell_i$ and $c(w_j) / |W(\ell_i)|$ is the estimate of the unigram
that completes the bigram~\cite{bahl83}. 

We further compensate the language model to account for future unseen 
bigrams by diverting some of the probability of the 
training bigrams to those that are as yet unobserved. 
We use the Modified Kneser-Ney (MKN) algorithm~\cite{kneser95}
for this compensation because it offers the best performance for 
interpolated language models~\cite{chen96}. The MKN algorithm 
subtracts a constant $\hat{d}$ from 
the observed frequency of every known bigram. It then estimates 
the likelihood that an unknown bigram $(w_{j-1},w_j)$ will appear with a 
modified estimate of the unigram $w_j$, where only the number 
of {\em distinct bigrams} that $w_j$ completes is considered:
$$P_c(w_j) = \frac{ |\{w: c(w,w_j) > 0\}|}{\sum_{v}|\{w: c(w,v) > 0\}|}$$
$P_c(w_j)$ is then weighted by the probability mass $\lambda(w_{j-1})$ 
that is taken by subtracting $\hat{d}$ from the counts of known bigrams: 
$$\lambda(w_{j-1}) = \frac{\hat{d}|\{w : c(w_{j-1},w) > 0\}|}{c(w_{j-1})}$$
Thus, under the MKN algorithm the probability of observing a bigram becomes:
\begin{equation}
  P_{\ell_i}(w_{j-1}, w_j) = \frac{\max(c(w_{j-1},w_j)-\hat{d},0)}{c(w_{j-1})}
    + \lambda(w_{j-1})P_c(w_j) \nonumber
\label{eq:two}
\end{equation}
If $(w_{j-1}, w_j)$ is unknown, the probability is just given 
by $\lambda(w_{j-1})P_c(w_j)$, and if it is known, the probability is
given as a linear interpolation of the modified bigram and unigram estimates. 
Note that the modified unigram estimate $P_c(w_j)$ is superior to 
$c(w_j) / |W(\ell)|$ because 
under $P_c(w_j)$, words that appear frequently but within few
distinct contexts will not strongly influence the probability of
the bigram. We estimate $\hat{d}$ such that the log-likelihood 
that the model generates a given bigram is maximized:
$$ \hat{d} = \arg\max_d \sum_v c(v,w_j) \log P_\ell(v,w_j)$$
This has a closed form approximation depending on whether 
$c(w_{i-1},w_i)$ is equal to $1$, $2$, or $\geq 3$~\cite{sundermeyer11}. 
Using these approximations, we set 
$\hat{d}$ equal to $d_1, d_2$, or $d_3$ respectively:
$d_1 = 1 - (2n_2/(n_1 + 2n_2))$, $d_2 = 2- (3n_3n_1/(n_2(n_1 + 2n_2)))$,
and $d_3 = 3-(4n_4n_1/(n_3(n_1 + 2n_2)))$
where $n_i$ is the number of bigrams that appear with frequency $i$.
Subsequently, we define the probability that a social media post
$T$ is generated from a sub-region $\ell_i$ $P(T | \ell_i)$ as: 
$$ P(T |\ell_i) = \prod_{j=2}^k P_{\ell_i}(w_{j-1}, w_j)$$

\subsection{Estimating Geo-Coordinates}
After training the language models over tweets 
from each sub-region, the ensemble is
queried to compute the probability that a social media post $T$ is 
generated from a sub-region $\ell_i$ using Bayes rule:
$$
P(\ell_i | T) = \frac{P(T | \ell_i)P(\ell_i)}{\sum_j P(T | \ell_j)P(\ell_j)}
$$
$P(\ell_i)$ is the prior probability that a social media post is 
from sub-region $\ell_i$ and is given by $N(\ell_i) / N(\mathcal{L})$.
$N(\ell_i)$ is the number of posts in $\ell_i$ and $N(\mathcal{L})$ is 
the total number of posts in the entire city $\mathcal{L}$. 
The geo-coordinates of a post $T$ may be estimated as the center of the 
sub-region whose posterior probability $P(\ell_i | T)$ is the highest,
that is, we may choose the center of $\ell^*$ 
where $\ell^* = \arg\max_i P(\ell_i | T)$.

Previous works suggest that the proximity to an object
increases the propensity of the users to post 
about it~\cite{dalvi12,serdyukov09}. 
In other words, it is feasible that the language of a sub-region may be 
influenced by the landmarks and events within its neighboring 
sub-regions. Thus, although we can naively use the highest $P(\ell_i | T)$ 
to estimate the geo-coordinates of a post, we introduce 
a {\em geo-smoothing} function $\Theta^\circ(\ell_i | T)$, which combines
the posterior probabilities of the neighboring sub-regions 
to capture their influence on the language in $\ell_i$.
Based on this geo-smoothing function, we select $\ell^*$ as 
$\ell^* = \arg\max_i \Theta^\circ(\ell_i | T)$. Popular functional 
forms for $\Theta^\circ$ include a decay component that reduces the 
contribution of neighbors as they get increasingly away from 
$\ell_i$~\cite{li12}. Such geo-smoothing performs best when the decay 
component takes a polynomial form~\cite{dalvi12,serdyukov09}.
Thus, in this preliminary study, we consider the simplest polynomial shown to 
be effective in geo-locating documents~\cite{serdyukov09}. Letting
$\Omega_k(\ell_i)$ be the set of neighbors of $\ell_i$ 
whose distance is $k$ cells away, and $P_{\ell_i}(T) = P(\ell_i | T)$,
our geo-smoothing function is defined as:  
\begin{align*}\begin{split}
&\Theta^\circ(P_{\ell_i}(T); \alpha, d) = (1-\alpha) P_{\ell_i}(T) \newln
	+\alpha\sum_{k=1}^d\sum_{\omega \in \Omega_k(\ell_i)} \frac{P_{\omega}(T)}{(2k+1)^2-1}
\end{split}\end{align*}
where $\alpha \in [0,1]$ is the smoothing weight and $d$ is
smoothing diameter, that is, the largest distance from which a neighbor
can be located. 

It is important to note that the accuracy of the estimated 
geo-coordinates is limited to the resolution of the $g \times g$ 
grid chosen to divide the region into sub-regions. 
Increasing $g$ will decrease the size of the sub-regions and allow 
for more accurate estimation, however, the number of posts 
available within each sub-region may be insufficient to 
train the models. On the other hand, decreasing $g$ increases the size 
of individual sub-regions so that they contain more posts, but limits 
the estimation accuracy. Similarly, increasing $\alpha$ and
$d$ respectively increase the importance and number of the 
neighboring sub-regions. If $\alpha$ and $d$ are very high, 
neighboring sub-regions may dwarf the candidate sub-region. 
However, if they are too low, they may not adequately capture users'
reactions on the local events and objects. We empirically choose the 
values of the three hyperparameters $g$, $\alpha$ and $d$ to balance 
these competing concerns. 

\section{Experimental Evaluation}
\label{sec:results}
In this section, we describe the data, its pre-processing, 
hyperparameter fits, and evaluation results. 

\subsection{Data Pre-Processing}
We collected over half-million geo-tagged tweets using Twitter's 
Streaming API~\footnote{\url{https://dev.twitter.com/docs/api/1.1/post/statuses/filter}} 
across New York City over a three-month period
(January 29th - April 7th, 2013). The tweets were 
collected across a $51.44$km$^2$ region that includes downtown and 
midtown Manhattan because it includes popular residential and
commercial districts as well as tourist destinations. We expect that
because of this diversity tweets from this region will capture 
varied thoughts representing the perspectives of 
long-term city residents, commuters, and visitors. 

For every tweet, we eliminated all non-English words 
and characters. We also eliminated hashtags because 
although they may indicate topic and 
content, they may also include shorthand or 
concatenated words (i.e. ``\#WestEnd'') that the language
models cannot decipher. We further pre-processed the tweets
by converting all words to lowercase and by stripping punctuation,
username replies, and links to Web pages. We also produced a stopword
list of the $200$ most frequently used words such 
as ``at'', ``the'', and ``or'', 
which lack contextual information, and hence, introduce noise 
into the estimation of bigrams. We choose a limited stopword
list that is approximately equal to $1\%$ of the number of distinct
words across the data collection region. We also include a ``catch all'' 
unigram ``$<$misc$>$'' to aggregate the probability of words that occur 
only once. This term thus accounts for the many miscellaneous, shorthand,
mis-spelled, and other user-specific notations that are uniquely common 
to Twitter. Of the 574,948 tweets, we reserved 408,095 ($70\%$) 
for training the language models, 83,990 ($15\%$) for evaluation,
and another 82,863 ($15\%$) as hold-out data for fitting the
hyperparameters. 

\subsection{Hyperparameter Fits}
We find values for the three hyperparameters, namely,
the grid size $g$, smoothing weight $\alpha$, and smoothing diameter
$d$ such that the average estimation error of
$\Theta^\circ$ across the set of hold-out tweets is 
minimized. We define the average estimation error 
as the geo-distance (in km) between the actual GPS coordinates of a tweet 
to the center of the sub-region that the model estimates it is from.
The parameters were fit via a 
standard grid search where $\alpha$ and $d$
were varied within their range of possible values, namely,
$\alpha \in \{0.1,0.2,...,1.0\}$ and $d \in \{1,...,g\}$.  
We chose to vary $g \in \{5,6,...,15\}$ because the estimation error 
increased when $g$ was outside this range
regardless of $\alpha$ and $d$. 

\begin{figure}[!ht]
\centering
\includegraphics[scale=0.55]{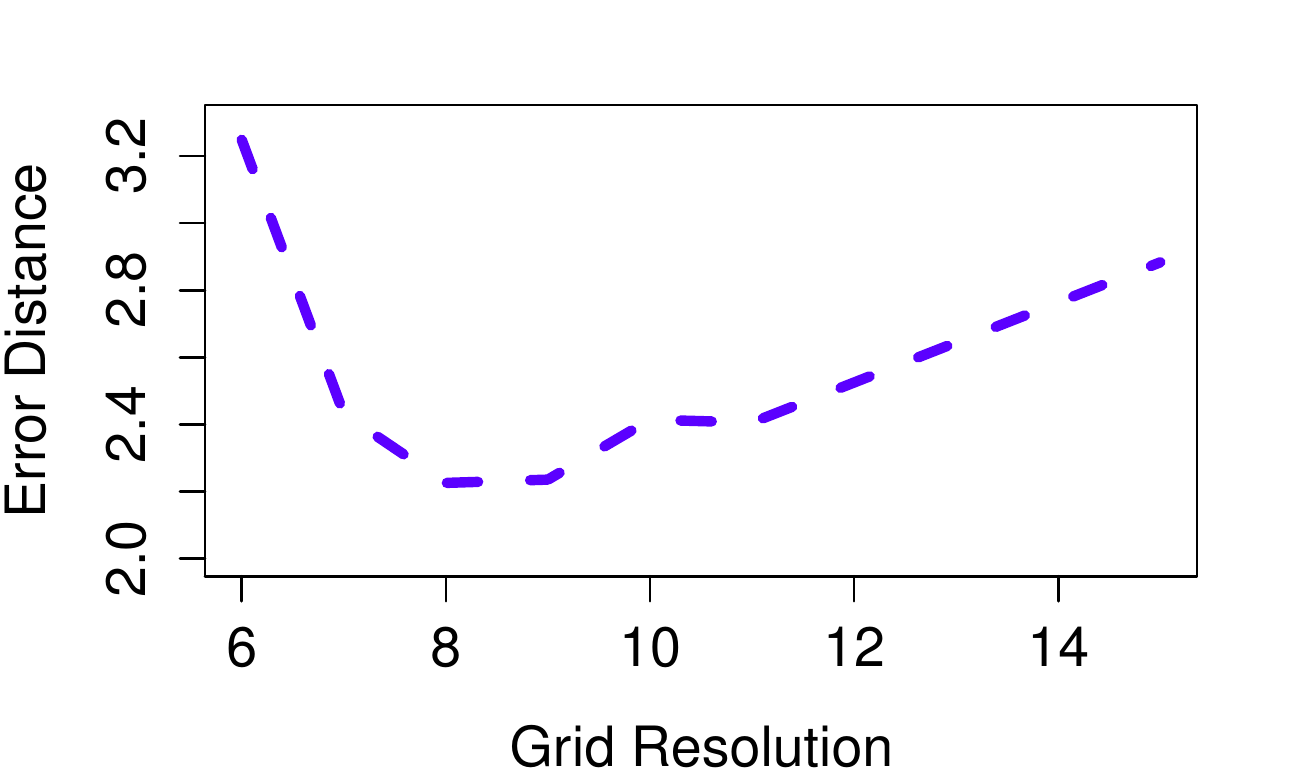}
\caption{Mean Hold-out Errors -- $\alpha=0.9$, $d=g$}
\label{fig:resolution}
\end{figure}

As we simultaneously varied $\alpha$, $d$, and $g$ in these ranges, 
we found that irrespective of $g$ and $d$, $\alpha = 0.9$ consistently 
minimized the estimation error. In other words, only $10\%$ of 
the posterior probability that a tweet $T$ originated from a 
sub-region $\ell_i$ can be attributed to its own language model 
$P_{\ell_i}(T)$, while the rest is contributed by the language 
models of the neighboring sub-regions. 
Furthermore, for every value of $g$, estimation error 
over the hold-out set is minimized at $d=g$. 
We thus set the grid size parameter $g$ equal to $d$  
and $\alpha = 0.9$. Figure~\ref{fig:resolution} shows the mean 
error for different values of $g$. We achieve the best performance
across the hold-out data when we set $g = 8$, where the
city is partitioned into 64 sub-regions each with an 
area $0.803$km$^2$. With $g$ set to $8$ and $\alpha=0.9$, 
we then estimate the 
the parameters of the interpolated bigram language models 
using the method described in Section~\ref{sec:method}.

\subsection{Evaluation Results}
We experimentally explore the influence of 
$d$ on the overall estimation error 
over the test set comprising 82,863 tweets. Table~\ref{tab:rep} shows 
that the average estimation error decreases as $d$ rises. 
Thus, even a simple polynomial function can appropriately decay 
probabilities from sub-regions as they get farther away from $\ell_i$ 
and significantly enhance estimation 
accuracy. We note that the mean error does not change for $d \geq 5$ 
because as the diameter of geo-smoothing increases
to include sub-regions $d=g/2$ cells away, central sub-regions in the city begin
to consider almost every other sub-region in its set of 
neighbors. Further increases in $d$ 
thus do not change $P_{\ell_i}(T)$ for a growing number
of sub-regions, causing the mean estimation error to converge. 

\begin{table}[!ht]
\begin{center}
\begin{tabular}{|c|c|c|c|c|c|} \hline   
   Diameter   &  $1$ & $2$ & $3$ & $4$ & $5+$ \\ \hline
   Mean Error & $3.38$ & $2.39$ & $2.18$ & $2.16$ & $2.15$ \\ \hline  
\end{tabular}
\caption{Mean Estimation Error (km)}
\label{tab:rep}
\end{center}
\end{table}

\begin{figure}[!ht]
\centering
\includegraphics[scale=0.55]{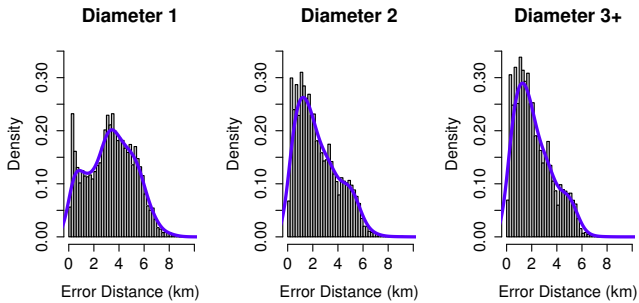}
\caption{Error Densities vs. $d$}
\label{fig:density}
\end{figure}

We evaluate the distribution of error estimates as a function of $d$ in
in Figure~\ref{fig:density}. At $d=1$, where only directly 
adjacent neighbors are considered, the estimation errors are bi-modal 
with small peaks at approximately $0.7$km and $3.8$km. The error 
distribution has a very wide variance; except for a decrease between $1$ and $3$km,
the error terms 
are generally distributed uniformly in the range $0$ and $6$km. 
The bi-modal behavior disappears for $d=2$ and most of the mass 
accumulates at errors less than than $2$km. 
For $d=3$, the peak sharpens even further at approximately $1.75$km,
which is less than the mean estimation error 
of $2.18$km. The densities for $d \geq 3$ are nearly identical 
because the estimates change only for a very small number of tweets 
as $d$ increases from $1$ to $3$.

\begin{figure}
\centering
\includegraphics[scale=0.5]{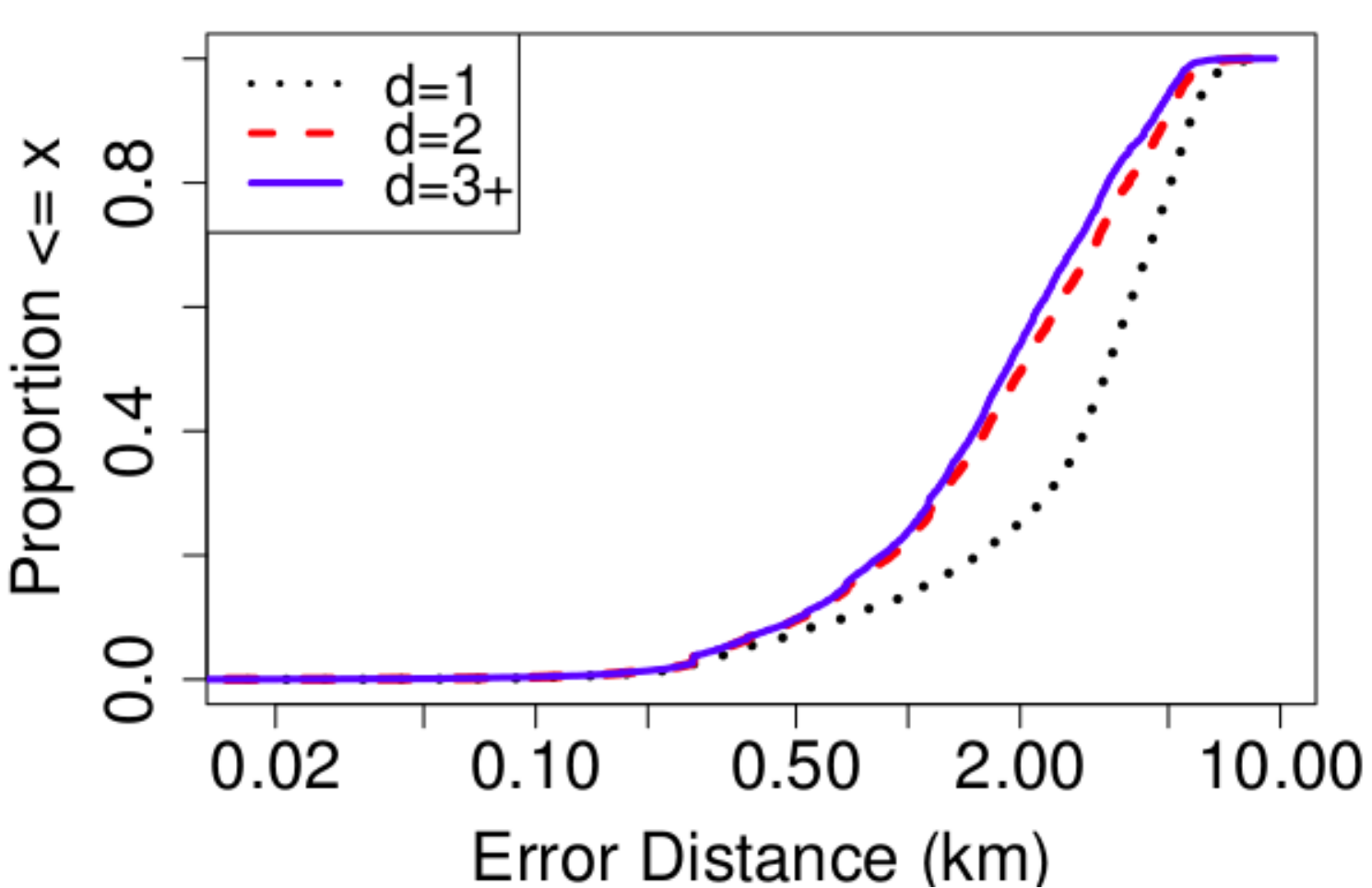}
\caption{Error Distributions}
\label{fig:cdf}
\end{figure}

Finally, we evaluate how frequently our model is accurate to within a 
given distance in
Figure~\ref{fig:cdf}, where we plot the CDF of estimation errors for
$d=1,2,$ and $\geq 3$ on semi-log scale. The shape of the distribution
function becomes linear for $d > 2$, which suggests that
the density function takes an exponential form as the smoothing diameter 
increases. We also find that the estimation accuracy increases 
only marginally beyond $d \geq 2$, suggesting
that smoothing over neighbors more than $2$ sub-regions away 
offers diminishing returns. Furthermore, the semi-log plots confirm
monotonic behavior. In other words, there is no special case or 
specific instance for which a smaller value of $d$ outperforms a 
larger value $d$. The overall accuracy of our approach is promising;
at $d=3$ the model can estimate the geo-coordinates of a 
tweet to within $4$km with probability $80\%$, to within $2$km
at over $50\%$, and to within $1$km at $20\%$. 

\section{Related Research}
\label{sec:rr}
In this section, we review contemporary techniques
according to whether they estimate the locations or geo-coordinates
of social medial posts. 

Broadly, the techniques that identify user locations either zero in on 
the ``home'' locations of
users or location ``types'' from where users check-in and update. 
These methods rely on spatial word 
usage and language models~\cite{dalvi12}, posting behaviors and 
external data on locations~\cite{mahmud12}, and 
inferences based on unified discriminative models~\cite{li12}. 
Estimation of specific geo-coordinates also use varied 
techniques including language models~\cite{serdyukov09,cheng10}, 
spatial word distributions~\cite{cheng10}, 
sequences of check-ins from social network friends~\cite{sadilek12}, 
integrating mobile phone data~\cite{cho11}, and mapping 
latent topics from across regions~\cite{hong12}.

Despite the integration of data from separate sources, and 
the aid of sophisticated probabilistic models, both types of 
approaches can identify only broad areas. By contrast, the 
methodology proposed in this paper can narrowly estimate geo-coordinates 
while relying solely on the content of the posts and on prior knowledge 
about the wide area from where they originated. 

\section{Conclusions \& Future Work}
\label{sec:conc}
In this paper, we presented a methodology to narrowly estimate
the geo-coordinates of a social media post, given the knowledge of
the much broader region from where they originate. An experimental
evaluation using tweets collected from New York
City shows that on an average the methodology can estimate geo-coordinates of 
social media posts to within $2.15$km, or just 4\% of the size of 
the broader region. Future work will examine the accuracy of
the approach over regions with distinct geographic features, 
sizes, and population distributions. We also propose to
investigate the accuracy of alternative geo-smoothing methods. 

\bibliographystyle{abbrv}
\bibliography{seke14}
\end{document}